\documentclass[sigconf,screen]{acmart} 

\startPage{1}
\setcopyright{none}
\usepackage{balance}
\usepackage{tabularx}
\usepackage{amsmath,amsfonts,bbm}
\usepackage{graphicx}
\usepackage{textcomp}
\usepackage{xcolor}
\usepackage{fancyhdr}
\usepackage{fancybox}
\usepackage{enumitem}
\usepackage{colortbl}
\usepackage{bigstrut}
\setlist{nosep} 
\usepackage{hyperref}
\usepackage{rotating}
\usepackage{booktabs}
\usepackage{balance}
\usepackage{algorithm}
\usepackage[noend]{algpseudocode}
\usepackage{multicol}
\usepackage{multirow}
\usepackage{soul}
\usepackage{blkarray}
\usepackage{lipsum,wrapfig}
\usepackage{caption}
\usepackage{subcaption}
\usepackage{tikz}
\usepackage{cleveref}
\usepackage{xspace}
\usepackage{algorithm}
\usepackage{algpseudocode}
\usetikzlibrary{shadows}
\usepackage{graphics}
\usepackage{array}
\usepackage{makecell}
\usepackage{mdframed}
\usepackage{boldline}
\usepackage{balance}
\usepackage[many]{tcolorbox} 


\AtBeginDocument{%
  \providecommand\BibTeX{{%
    \normalfont B\kern-0.5em{\scshape i\kern-0.25em b}\kern-0.8em\TeX}}}

\copyrightyear{2022}
\acmYear{2022}
\setcopyright{acmcopyright}\acmConference[ASE '22]{37th IEEE/ACM International Conference on Automated Software Engineering}{October 10--14, 2022}{Rochester, MI, USA}
\acmBooktitle{37th IEEE/ACM International Conference on Automated Software Engineering (ASE '22), October 10--14, 2022, Rochester, MI, USA}
\acmPrice{15.00}
\acmDOI{10.1145/3551349.3556960}
\acmISBN{978-1-4503-9475-8/22/10}

\begin{document}

\sethlcolor{lightgray}
\definecolor{pink}{RGB}{252,145,149}
\definecolor{lightpink}{RGB}{252,145,149}
\definecolor{lightgray}{gray}{0.8}
\definecolor{darkgray}{gray}{0.6}
\definecolor{Gray}{rgb}{0.88,1,1}
\definecolor{Gray}{gray}{0.85}
\definecolor{Blue}{RGB}{0,29,193}
\definecolor{babyblue}{RGB}{137, 207, 240}
\definecolor{MyDarkBlue}{rgb}{0,0.08,0.45} 
\definecolor{pink}{RGB}{231,95,110}
\definecolor{greenish}{RGB}{182, 231, 142}
\definecolor{orangish}{RGB}{255, 206, 144}
\definecolor{lavender}{RGB}{225, 213, 231}
\definecolor{purple}{RGB}{131, 84, 141}
\definecolor{army}{RGB}{113, 123,104}
\definecolor{lightergray}{rgb}{0.85, 0.85, 0.85}
\definecolor{lightestgray}{rgb}{0.95, 0.95, 0.95}
\definecolor{codebg}{HTML}{F4F4F4}
\definecolor{blueish}{RGB}{177, 206, 232}
\definecolor{redish}{RGB}{233, 86, 60}
\newcommand{\red}[1]{{\color{redish}#1}\xspace}
\newcommand{\Red}[1]{\textbf{\color{redish}#1}\xspace}
\definecolor{gray05}{gray}{0.95}
\definecolor{gray10}{gray}{0.90}
\definecolor{gray15}{gray}{0.85}
\definecolor{gray20}{gray}{0.80}
\definecolor{gray25}{gray}{0.75}
\definecolor{gray30}{gray}{0.70}
\definecolor{gray35}{gray}{0.65}
\definecolor{gray40}{gray}{0.60}
\definecolor{gray45}{gray}{0.55}
\definecolor{gray50}{gray}{0.50}
\definecolor{gray55}{gray}{0.45}
\definecolor{gray60}{gray}{0.40}
\definecolor{gray65}{gray}{0.35}
\definecolor{gray70}{gray}{0.30}
\definecolor{gray75}{gray}{0.25}
\definecolor{gray80}{gray}{0.20}
\definecolor{gray85}{gray}{0.15}
\definecolor{gray90}{gray}{0.10}
\definecolor{gray95}{gray}{0.05}

\newcommand{\bi}{\begin{itemize}[leftmargin=*, wide=0pt]}
\newcommand{\ei}{\end{itemize}}
\newcommand{\beq}{\begin{equation}}
\newcommand{\eeq}{\end{equation}}
\newcommand{\be}{\begin{enumerate}[leftmargin=*, wide=0pt]}
\newcommand{\ee}{\end{enumerate}}

\crefformat{section}{\S#2#1#3} 
\crefformat{subsection}{\S#2#1#3}
\crefformat{subsubsection}{\S#2#1#3}
\newcommand{\tion}[1]{\cref{sect:#1}\xspace}
\newcommand{\fig}[1]{Fig.~\ref{fig:#1}\xspace}
\newcommand{\tab}[1]{Table~\ref{fig:#1}\xspace}
\newcommand{\algo}[1]{Listing~\ref{fig:#1}\xspace}

\newcommand\CARGO{CARGO\xspace}
\newcommand\ie{i.e.\xspace}
\newcommand\eg{e.g.\xspace}


\newmdenv[
    tikzsetting= {fill=blue!10},
    skipabove=0.33em,
    skipbelow=0.33em,
    linewidth=1pt,
    innerleftmargin=4pt,
    innerrightmargin=4pt,
    innertopmargin=2pt,
    innerbottommargin=2pt,
    linecolor=gray95,
    roundcorner=2pt, 
    shadow=true,
    shadowsize=2pt,
    shadowcolor=black
]{myshadowbox}
\newcommand*\circled[1]{\tikz[baseline=(char.base)]{
            \node[shape=circle,fill,inner sep=2pt] (char) {\footnotesize\textcolor{white}{\bfseries #1}};}}
\newcommand*\circleda{\tikz[baseline=(char.base)]{
            \node[shape=circle,fill=purple,inner sep=2pt] (char) {\footnotesize\textcolor{white}{\bfseries A\xspace}};}}
\newcommand*\circledb{\tikz[baseline=(char.base)]{
            \node[shape=circle,fill=army,inner sep=2pt] (char) {\footnotesize\textcolor{white}{\bfseries B\xspace}};}}
\newcommand*\circledw[1]{\tikz[baseline=(char.base)]{
            \node[shape=circle,draw,inner sep=2pt] (char) {\footnotesize #1};}}
\renewcommand\tabularxcolumn[1]{m{#1}} 
\newcolumntype{s}{>{\centering \arraybackslash \hsize=.5\hsize}X}  %

\newenvironment{result}
{\centering
\begin{tabular}{p{0.95\linewidth}}  
  \cellcolor{gray!15}\textbf{Summary:}~}
{\\\end{tabular}}



\newcommand{\squishlist}{
 \begin{list}{$\bullet$}
 { \setlength{\itemsep}{0pt}
   \setlength{\parsep}{1pt}
   \setlength{\topsep}{1pt}
   \setlength{\partopsep}{0pt}
   \setlength{\leftmargin}{1em}
   \setlength{\labelwidth}{1em}
   \setlength{\labelsep}{0.5em} } }

\newcommand{\squishlisttwo}{
 \begin{list}{$\bullet$}
 { \setlength{\itemsep}{0pt}
  \setlength{\parsep}{0pt}
  \setlength{\topsep}{0pt}
  \setlength{\partopsep}{0pt}
  \setlength{\leftmargin}{0em}
  \setlength{\labelwidth}{1em}
  \setlength{\labelsep}{0.5em} } }

\newcommand{\squishend}{
 \end{list} }

\newcommand{\reminder}[1]{\textcolor{red}{[[ #1 ]]}\typeout{#1}}
\newcommand{\reminderR}[1]{\textcolor{gray}{[[ #1 ]]}\typeout{#1}}
\newcommand\rayb[1]{\textcolor{red}{{\bfseries RAYB:~#1\xspace}}}
\newcommand\rahul[1]{\textcolor{blue}{{\bfseries RAHUL:~#1\xspace}}}
\newcommand\rahulR[1]{\textcolor{babyblue}{{\sout{\bfseries RAHUL:~#1\xspace}}}}



\def\rot{\rotatebox}

\newcommand\cb{\cellcolor{blue!10}}
\newcommand\cg{\cellcolor{greenish!20}}
\newcommand\cp{\cellcolor{orange!20}}

\newcolumntype{C}[1]{>{\centering\let\newline\\\arraybackslash\hspace{0pt}}m{#1}}

\title{\CARGO: AI-Guided Dependency Analysis for Migrating Monolithic Applications to Microservices Architecture}

\author{Vikram Nitin}
\authornote{This work was done when the author was an intern at IBM Research.}
\email{vikram.nitin@columbia.edu}
\affiliation{%
  \institution{Columbia University}
  \country{New York, USA}}

\author{Shubhi Asthana}
\email{sasthan@us.ibm.com}
\affiliation{%
  \institution{IBM Research}
  \country{California, USA}
}

\author{Baishakhi Ray}
\email{rayb@cs.columbia.edu}
\affiliation{%
  \institution{Columbia University}
  \country{New York, USA}}

\author{Rahul Krishna}
\authornote{Corresponding author.}
\email{rkrsn@ibm.com}
\affiliation{%
  \institution{IBM Research}
  \country{New York, USA}}

\renewcommand{\shortauthors}{Nitin, V., et al.}

\begin{abstract}
Microservices Architecture (MSA) has become a de-facto standard for designing cloud-native enterprise applications due to its efficient infrastructure setup, service availability, elastic scalability, dependability, and better security.
Existing (monolithic) systems must be decomposed into microservices to harness these characteristics.
Since manual decomposition of large scale applications can be laborious and error-prone, AI-based systems to decompose applications are gaining popularity. However, the usefulness of these approaches is limited by the expressiveness of the program representation
and their inability to model the application's dependency on critical external resources such as databases. Consequently, partitioning recommendations offered by current tools result in architectures that result in (a)~distributed monoliths, and/or (b)~force the use of (often criticized) distributed transactions.
This work attempts to overcome these challenges by introducing CARGO~({short for \underline{\textbf{C}}ontext-sensitive l\textbf{\underline{A}}bel p\underline{\textbf{R}}opa\underline{\textbf{G}}ati\underline{\textbf{O}}n})—a novel un-/semi-supervised partition refinement technique that uses a context- and flow-sensitive system dependency graph of the monolithic application
to refine and thereby enrich the partitioning quality of the current state-of-the-art algorithms. CARGO was used to augment four state-of-the-art microservice partitioning techniques (comprised of 1 industrial tool and 3 open-source projects). These were applied on five Java EE applications (comprised of 1 proprietary and 4 open source projects). Experiments show that CARGO is capable of improving the partition quality of all four partitioning techniques. Further, CARGO substantially reduces distributed transactions, and a real-world performance evaluation of a benchmark application (deployed under varying loads) shows that CARGO also lowers the overall the latency of the deployed microservice application by 11\% and increases throughput by 120\% on average. 
\end{abstract}


\maketitle

\section{Introduction}

Businesses are gradually migrating their existing applications to the cloud as their enterprise applications outgrow their monolithic architectures. This allows them to leverage better scalability and reliability, faster development, easier maintenance and deployment, and better fault isolation, among others \cite{thones2015microservices}.
This migration requires a decomposition of monolithic applications into loosely connected compositions of specialized microservices \cite{richardson2018microservices,newman2021building,fowler_2004}. However, modernizing an enterprise application with entrenched technology stacks is a challenging task. Manual approaches to modernization consume a significant amount of engineering time and resources~\cite{github_rails} and a complete rebuild is rarely feasible. Therefore, big enterprises such as IBM \cite{kalia2021mono2micro}, Amazon \cite{amazon_m2m}, and GitHub \cite{github_rails} frequently advocate for partitioning applications by identifying functional boundaries in the code that may be extracted as microservices.
This has led to a rapid growth of research in using program analysis to  \textit{automatically} discover these functional boundaries and partitions within the application \cite{ahmadvand2016requirements, baresi2017microservices, chen2017monolith, escobar2016towards, gysel2016service, hassan2017microservice, klock2017workload, levcovitz2016towards, mazlami2017extraction}.


\noindent\textbf{Existing work and limitations.}~A typical microservice extraction approach starts by converting a monolithic application into a call-graph or control-flow graph representation. Then, an off-the-shelf graph clustering algorithm is used to find loosely connected functional boundaries and make partitioning recommendations. The performance of these graph clustering algorithms is directly influenced by the expressiveness and completeness of the graphs they use. Unfortunately, the program graphs used by current techniques are ill-suited to adequately model enterprise applications. The key limitations of current approaches include: 
\bi
    \item \textit{Missing database interactions:} Existing AI-guided microservice partitioning methods miss the interconnections between applications and databases. Transactional databases, which are stored and maintained centrally, are typically used by monolithic enterprise applications to execute and persist key business logic. If these interactions are not taken into account, the partitioning guidelines could force the adoption of distributed transactions (such as 2 Phase commits). This represents a considerable effort for developer who must then ensure that state changes are coordinated across several services. Further, distributed transactions pose other challenges such as: a) lack of consistency in the data; b) probable safety violations; and c) loss of ACID isolation guarantees \cite{hohpe2002enterprise,ruecker2021practical,kleppmann2017designing}.
    \item \textit{Imprecise program analysis:} Popular approaches such as CoGCN \cite{Desai+AAAI+2021} and microservice extractor \footnote{https://aws.amazon.com/microservice-extractor/} use context-insensitive static-analysis which tends to be incomplete and imprecise. Further, these techniques use off-the-shelf static analysis tools such as SOOT \cite{lam2011soot} and WALA \footnote{https://github.com/wala/WALA} which are known to ignore crucial features used by enterprise java applications such as dependency injection \cite{fowler2006inversion}, and centralized object store \cite{johnson2004expert}, among others. This often results in a sparse coverage of several critical sections of the application \cite{antoniadis2020static}. To overcome this challenge, certain approaches construct a dynamic call graph by instrumenting the program to amass light-weight traces as the application's many use-cases are exercised~\cite{kalia2021mono2micro,Jin+TSE+2019}. However, for dynamic analysis to be useful, it must be run with sufficient inputs to expose as many different program behaviors as possible, and this can be hard to satisfy. Further, building call-graphs with dynamic tracing has additional drawbacks---it can: (a)~be intrusive, (b)~lead to considerable performance degradation, and (c)~produce an incomplete representation of the application.
\ei

\noindent\textbf{Our approach.}~To address the above challenges, we propose a microservice partitioning and refinement tool that statically analyzes JEE applications to build a comprehensive, highly-precise, context-sensitive system dependency graph (SDG). Our SDG is a rich graphical abstraction of a Java EE application with various crucial relationships that exist in the application, namely: call-return edges, data-flow edges, heap-dependency edges, and database transaction edges. Our work also proposes a novel community detection algorithm called \underline{\textbf{C}}ontext-sensitive l\textbf{\underline{A}}bel  p\underline{\textbf{R}}opa\underline{\textbf{G}}ati\underline{\textbf{O}}n (abbreviated as \CARGO, serving as an acronym for our general tool) that isolates ``snapshots'' of the system dependency graph under various contexts to discover and/or to refine existing partitions into \textit{highly-cohesive} and \textit{loosely-coupled} compositions of the program which can be implemented as microservices.


\noindent\textbf{Results summary.}
We evaluate \CARGO on 5 monolithic applications (4 open source and 1 proprietary) a suite of architectural metrics. Further, using \CARGO's partitions, we modernize a benchmark monolithic application (daytrader) to evaluate its real-world performance in terms of latency and throughput. Our key findings are listed below:
\bi
    \item \CARGO is able to almost eliminate distributed database transactions, both when applied in conjunction with a baseline approach and when applied stand-alone.
    \item We use \CARGO to refine the partitions produced by a baseline approach, and find that the resulting microservice application has 11\% less latency and 120\% more throughput than the baseline.
    \item When we compare \CARGO to baseline partitioning approaches on 4 architectural metrics, \CARGO is better than the baselines on 3 metrics and inferior on 1 metric.
\ei

\noindent\textbf{Contributions.}~\CARGO makes the following contributions to improve the current state of the art:
\be
\item 
\CARGO builds a highly precise statically derived system dependency graph (SDG) with full context and flow sensitivity (detailed in \cref{sect:methods}). The nodes of the SDG represent the methods and database tables in the application. The various inter-procedural edges express different types of program dependencies between one method and another, and transactional edges highlight relationships between methods and database tables. Together, the SDG constructed by \CARGO offers a \textit{fine-granular view} of an application that includes all its internal and external (\ie, database) relationships.
\item 
\CARGO introduces a novel partitioning algorithm based on label propagation (aka. LPA)~\cite{xiaojin2002learning} called \textit{context-sensitive label propagation}. Context-sensitive label propagation (detailed in \tion{methods}) extracts ``context-snapshots'' of the SDG, \ie abstractions of the possible \textit{dynamic states} of the program~\cite{milanova2005parameterized}, and applies label propagation across all contexts to identify the best functional boundaries in code.
\item \CARGO offers multiple modes of usage: (i)~as a stand-alone tool, wherein the partitions are discovered in an unsupervised manner; or (ii)~as an add-on to existing tools, wherein the existing partitions are refined to improve their overall quality. 
\item \CARGO and the associated artifacts will soon be made publicly available on Github as part of the Konveyor Tackle DGI project. \footnote{https://github.com/konveyor/tackle-data-gravity-insights}
\ee

\section{Background and Motivation}
\label{sect:motivation}
A recent report shows that only 20\% of the enterprise workloads are in the cloud, and they were predominately written for native cloud architectures. This leaves 80\% of applications as \textit{monoliths} deployed on-premises, waiting to be refactored  and modernized for the cloud. In a monolithic architecture the entire application is bundled into a single package in a monolithic architecture, which includes source code, libraries, configurations, and any other dependencies required for it to execute. Monoliths aren't intrinsically bad, although they do have several drawbacks: (a) Deployment artifacts are often huge, sluggish to start, and consume a substantial amount of resources; (b) changing one component of the program necessitates rebuilding and redeploying the entire application, which slows developer productivity---a challenge that triggered GitHub to migrate away from monoliths to a microservice architecture\footnote{https://www.infoq.com/articles/github-monolith-microservices/}.

In comparison to monoliths, microservices split an application into discrete units that have clearly defined service boundaries. This offers a number of benefits such as: continuous delivery of large and complex applications, small and easily maintainable applications, low resource-intensiveness, independently deployable and scalable applications, easy fault isolation, and autonomy in development. 

The migration from monolith to a microservice architecture is no mean feat. Even a small, seemingly trivial, upgrade to a monolithic software can be time consuming and resource intensive. For example, a rails upgrade to a 10 year old system at GitHub took a year and a half to complete~\cite{github_rails}. Therefore, the best strategy to migrate monolithic applications to microservices is to do so incrementally by identifying boundaries in the application such that the functionalities encompassed by them are highly cohesive, yet as loosely coupled as possible to other functionalities in the code.

The study of identifying these boundaries has seen a lot of interest lately with prominent industry tools such as Mono2Micro\footnote{https://www.ibm.com/cloud/blog/announcements/ibm-mono2micro} and several others in academia~\cite{ahmadvand2016requirements, baresi2017microservices, chen2017monolith, escobar2016towards, gysel2016service, hassan2017microservice, klock2017workload, levcovitz2016towards, mazlami2017extraction, kalia2021mono2micro}. However, when confronted with enterprise applications, each of these approaches faces some difficulties. We will look at a common motivating scenario below and the rest of this paper attempts to address this problem. 

\subsection{Motivation}

In this section, we use a real-world example to highlight the challenges that may arise when attempting to partitioning a monolith. We use a state-of-the-art microservice partitioning algorithm to partition a monolithic version of a popular JEE trading application called daytrader\footnote{\url{https://github.com/WASdev/sample.daytrader7}}. \fig{motivation} shows an example with three classes from daytrader: \texttt{QuoteDatabean}, \texttt{TradeDirect}, and \texttt{TradeConfig}. \fig{motivation_a} shows the most common partitioning recommendation of these three classes where, partition \circleda~~contains \texttt{QuoteDatabean} and partition \circledb~~contains \texttt{TradeDirect} and \texttt{TradeConfig}.

At first glance, this may seem like a reasonable partitioning strategy (especially when inspecting the call-edges between the classes in~\fig{motivation_a}), but a closer look (as seen in~\fig{motivation_b}) paints a very different picture. Several challenges will arise if a developer tried to partition according to the recommendations made by current approaches, as in \fig{motivation_a}. Some of these are described below:

\begin{figure}[t!]
    \centering
    \begin{subfigure}{\linewidth}
    \includegraphics[width=0.95\linewidth]{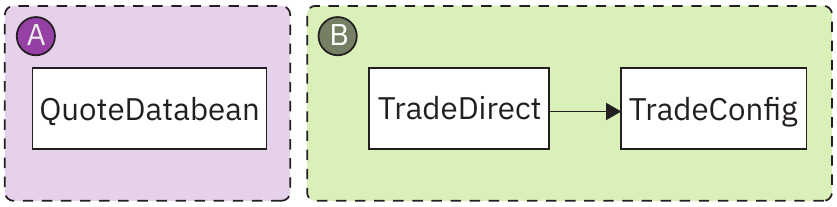}
    \caption{\small A typical partitioning recommendation for two important classes in Daytrader offered by FoSCI and CoGCN.}
    \label{fig:motivation_a}
    \end{subfigure}
    \par\bigskip
    \begin{subfigure}{\linewidth}
    \includegraphics[width=0.95\linewidth]{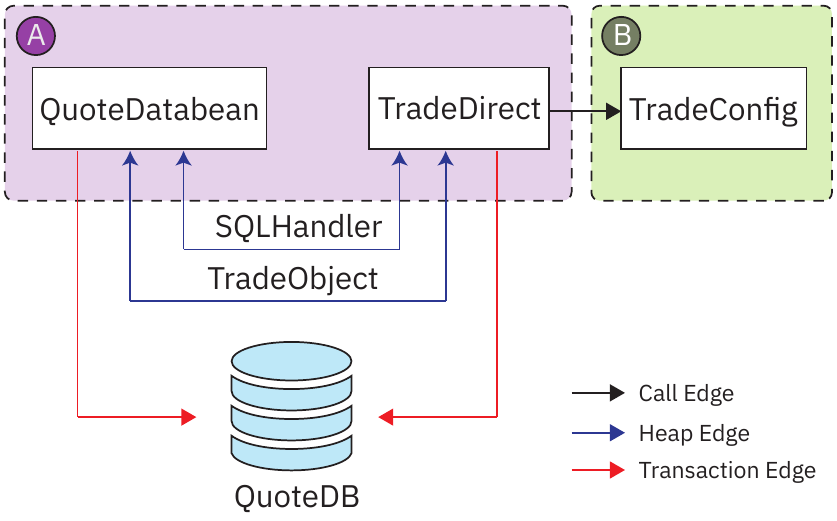}
    \caption{Recommended partitioning considering the the high coupling between \texttt{QuoteDataBean} and \texttt{TradeDirect}.}
    \label{fig:motivation_b}
    \end{subfigure}
    \caption{\small An example from Daytrader that illustrates a distributed monolithic behavior. The two classes \colorbox{lavender}{\texttt{QuoteDatabean}} and \colorbox{greenish!50}{\texttt{TradeDirect}} lie in different partitions, yet they are tightly coupled by their: a)~shared references to \texttt{TradeObject} and \texttt{SQLConnector}; and b)~part-taking in a distributed transaction with \colorbox{babyblue!60}{QuoteDB}.}
    \label{fig:motivation}
\end{figure}

\noindent \textbf{Challenge \#1:~}\textit{Services are tightly coupled leading to distributed monolithic behavior.}~Coupling refers to the degree of separability between two classes. There exists a tight coupling between the \texttt{QuoteDatabean} and \texttt{TradeDirect}, although not apparent in \fig{motivation_a}, due to heap-dependencies induced by \texttt{SQLHandler} and \texttt{TradeObject}. These objects carry the payload of information exchange between the two classes that is crucial to execute the business functionality. If partitioned according to \fig{motivation_a}, one would have to rely on serialization/de-serialization protocols as well as tightly choreographed object state management to ensure the business logic is executed accurately. This makes the partition behave more like a monolith that is deployed as microservices (aka. a distributed monolith) rather than a true microservice application.

\noindent \textbf{Challenge \#2:~}\textit{Services engage in transactions to the same database table.}~If partitioned according to~\fig{motivation_a}, we would be forced to reconcile a distributed transaction between \texttt{QuoteDatabean}, \texttt{TradeDirect}, and the database table \texttt{QuoteDB} (as is seen in~\fig{motivation_b}). This would require a developer to implement 2 Phase commits (or 2PCs) to coordinate the state change across \texttt{QuoteDatabean}, \texttt{TradeDirect}. The use of 2PCs can be very problematic due a number of reasons including but not limited to (1)~Loss of atomicity and/or isolation guarantees, (2)~Non-trivial coordination of distributed locks, and (3)~Very high latency. Therefore they must be avoided if possible lest one is forced to rely on expensive data centers and meticulously synchronized atomic clocks~\footnote{\url{https://www.youtube.com/watch?v=iKQhPwbzzxU}}. 

We note that these challenges are very common, and appear often in enterprise scale systems. As a consequence, the recommendations from certain microservice partitioning tools are often rendered useless due to the potential ramifications of implementing their recommendations. This inspired the construction of \CARGO, where we attempt to refine the recommendations of existing approaches to reduce incidents of distributed monoliths and/or distributed transactions. The partitions in~\fig{motivation_b} were derived from \CARGO. Here, \texttt{QuoteDatabean} and \texttt{TradeDirect} have been grouped in the same partition, essentially overcoming the above challenges.

\begin{figure*}[t!]
    \centering
    \includegraphics[width=\linewidth]{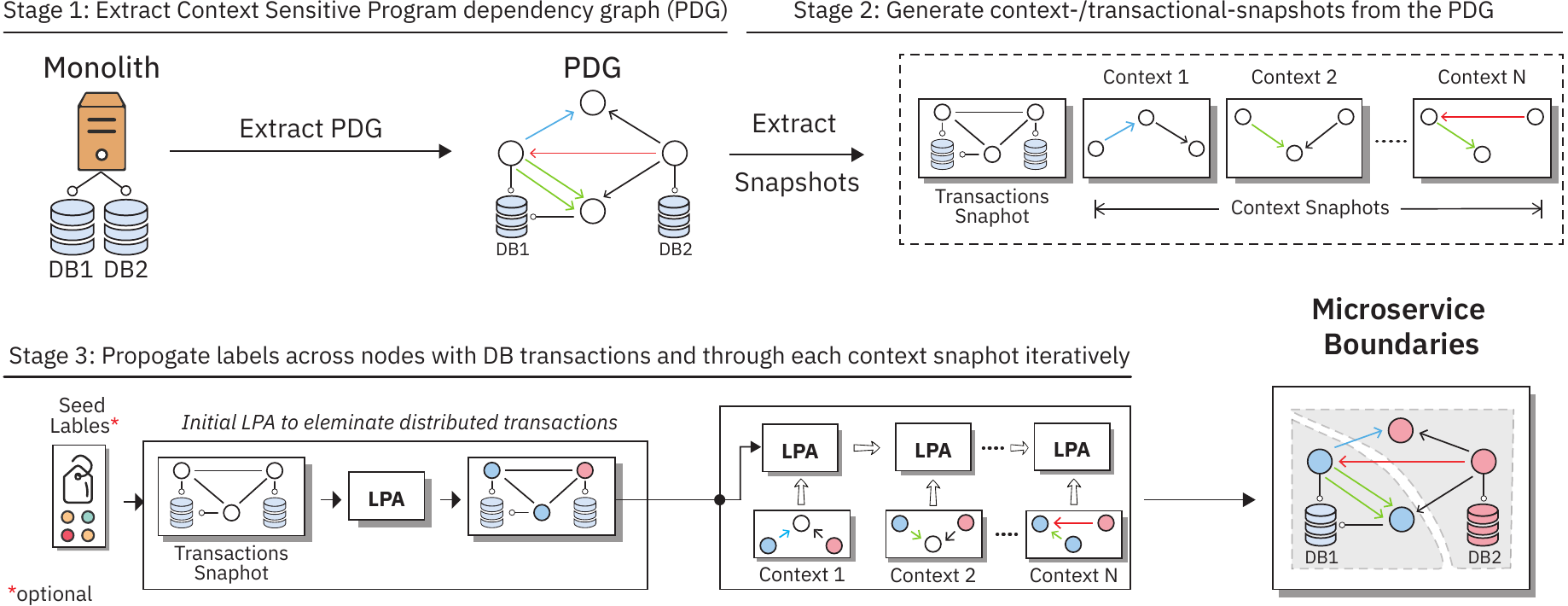}
    \caption{Overview of CARGO.}
    \label{fig:cargo_overview}
\end{figure*}

\section{Methodology}
\label{sect:methods}

This section describes \CARGO and how it identifies (and/or refines) microservice partitions in monolithic enterprise Java applications to obtain loose coupling between services and high cohesion within the services.
~\fig{cargo_overview} illustrates an end-to-end workflow of the proposed approach. In the first stage (\tion{stage1}), we build a context and flow sensitive system dependence graph (SDG) of an enterprise application which has methods and DB tables as nodes with various edges between them such as call-return edges, data-dependency edges, heap-dependency edges, and transactional edges. In the second stage (\tion{stage2}), we extract sub-graphs of the SDG under each context and across all the transactional scopes. We refer to these as \textit{contextual snapshots}. In the final stage (\tion{cargo_algo}), we apply a novel label propagation algorithm across each contextual snapshot to group functionally similar components of the program together.

\subsection{Building a System Dependency Graph}
\label{sect:stage1}

In this section, we describe the construction of a System Dependence Graph (SDG)~\cite{ferrante1987program, horwitz1990interprocedural}, which is an instrumental component of \CARGO. The SDG captures the semantics of the whole program and models different program contexts explicitly.
%
%
%
This makes it particularly well-suited to model large enterprise applications, since context-sensitive analyses offer a precise way to distinguish different invocations of the application without need for expensive dynamic tracing. The rest of this section first describes context-sensitivity followed by the various dependency (edge) types in the SDG (\tion{dependencies}).

\subsubsection{Context sensitivity}
\label{sect:context-sen}
Context-sensitivity dictates that every relationship in the SDG is qualified by a context. This ensures that different calls/information flows across a method are annotated with a unique \textit{identifier}. This enables us to distinguish between different paths through the same section of the code, thereby enhancing the precision and richness of our SDG. 

Consider the sample program in~\fig{context_sen}: here, in lines 2 and 5, we allocate two objects \texttt{a1} and \texttt{a2} of type \texttt{A}, and for each of these objects, we call the class method \texttt{A.foo($\ldots$)} which in turn calls \texttt{B.bar($\ldots$)} (on line 12) and \texttt{C.id($\ldots$)} (on line 19).
\begin{figure}[htb!]
\centering
\includegraphics[width=\linewidth]{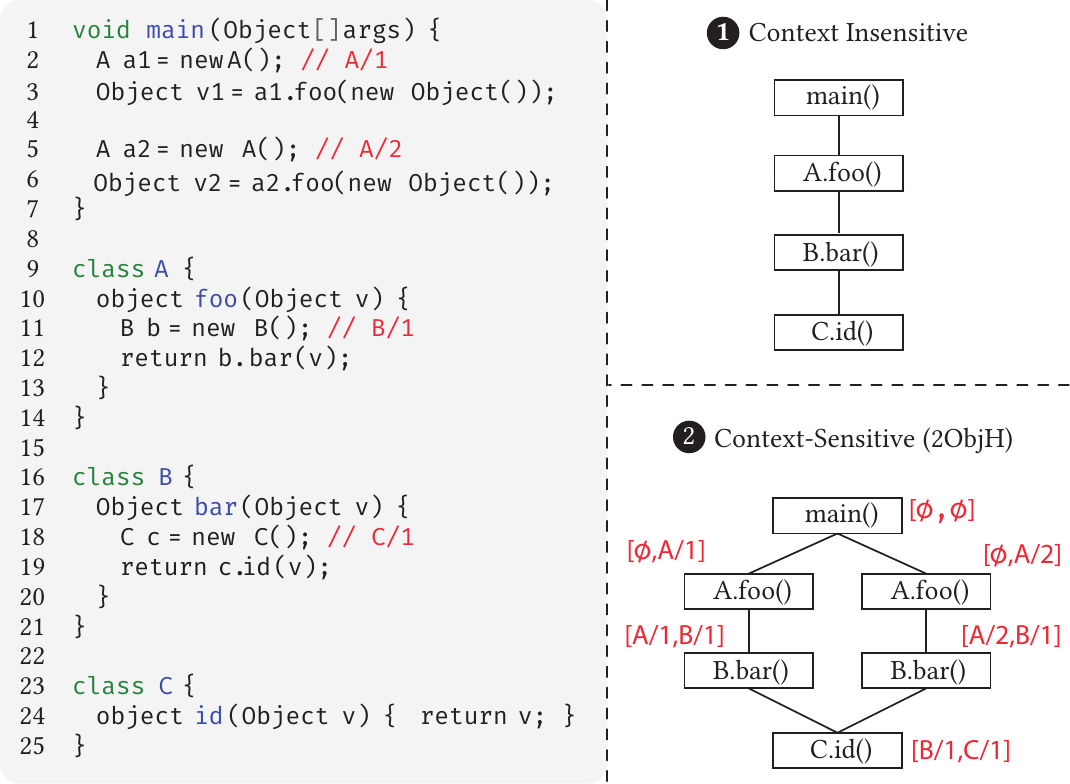}
\caption{Difference between a context-sensitive and a context insensitive call graph. Context-insensitive analysis cannot differentiate between two calls to the same method, context-sensitive analysis uses object allocation sites (highlighted in \red{red}) to distinguish between calls.}
\label{fig:context_sen}
\end{figure}
A context-insensitive view of this program as seen in {\footnotesize \circled{1}}, fails to distinguish between two \textit{distinct} calls to \texttt{B.bar($\ldots$)}. Whereas, a context-sensitive analysis as seen in  {\footnotesize\circled{2}} distinguishes between the call arising from \texttt{a1.foo($\ldots$)} and \texttt{a2.foo($\ldots$)} as separate edges.

Contexts may be thought of as abstractions of the possible dynamic executions of a program, \ie, a method and its associated variables under two different contexts represent non-overlapping program states that may be reached in one or more dynamic runs of the program. The main flavors of context sensitivity are call-site-sensitivity~\cite{sharir1978two,shivers1991control}, object-sensitivity~\cite{milanova2005parameterized}, and type-sensitivity~\cite{smaragdakis2014introspective}. Of these, object-sensitivity has been shown to be remarkably precise for analyzing object-oriented programs like Java~\cite{smaragdakis2014introspective,jackee}. Contexts are qualified using the object allocation site the current method was invoked on, \eg, in \fig{context_sen} contexts are qualified at \texttt{\red{A/1}}, \texttt{\red{A/2}}, \texttt{\red{B/1}}, and \texttt{\red{C/1}}. Thus, the contexts of two invocations may differ even if their call-site was the same but the object that the method was called on was different. For example, in {\footnotesize\circled{2}}, \texttt{B.bar($\ldots$)} appears in two contexts, even though it has the same call-site (on line 11).

In this paper, we use 2-object-sensitive-analysis with a context-sensitive heap (2objH) for all our analysis as this has been considered the gold-standard for scalable yet precise analysis~\cite{jackee,li2018precision,li2018scalability}. Although even more precise analyses are available, and can be developed, they seldom scale when applied to enterprise JAVA  applications~\cite{jackee}~\footnote{A detailed discussion of context-sensitivity is beyond the scope of this work; we therefore direct interested readers to some influential work in this area~\cite{jackee,smaragdakis2011pick,milanova2005parameterized}.}. A crucial property of the object-sensitivity analysis is the \textit{inherent flow-sensitivity} imparted by the contexts. As illustrated in~\fig{context_sen}, each object allocation site is annotated with a numeric identifier (\eg, \texttt{\red{A/1}}, \texttt{\red{A/2}}). These identifiers indicate the flow of the program, with \texttt{\red{A/1}} and all its subsequent contexts appearing before \texttt{\red{A/2}} and all the contexts that follow. This facilitates accurate reasoning about potential control-/data-flows in the program.

We note that context sensitivity is not limited to just call-graphs; it may be used for any static analysis such as data-flow analysis, alias analysis, heap-dependency analysis, and more.

\subsubsection{Dependency (edge) types} 
\label{sect:dependencies}

\begin{figure}
\centering
\begin{subfigure}{\linewidth}
\includegraphics[width=\linewidth]{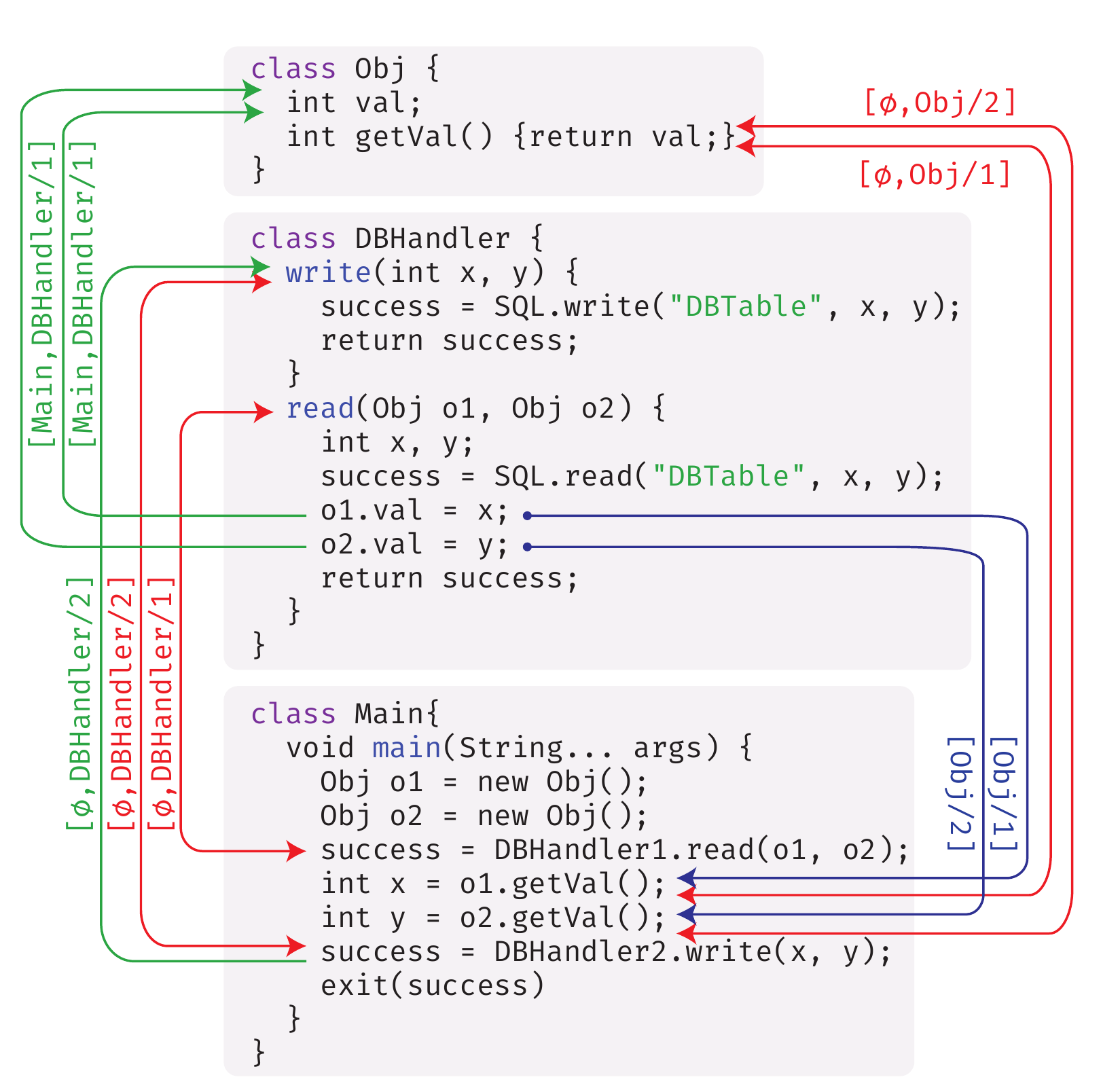}
\caption{An example program}
\label{fig:dependencies_a}
\end{subfigure}
\\
\begin{subfigure}{\linewidth}
\includegraphics[width=0.9\linewidth]{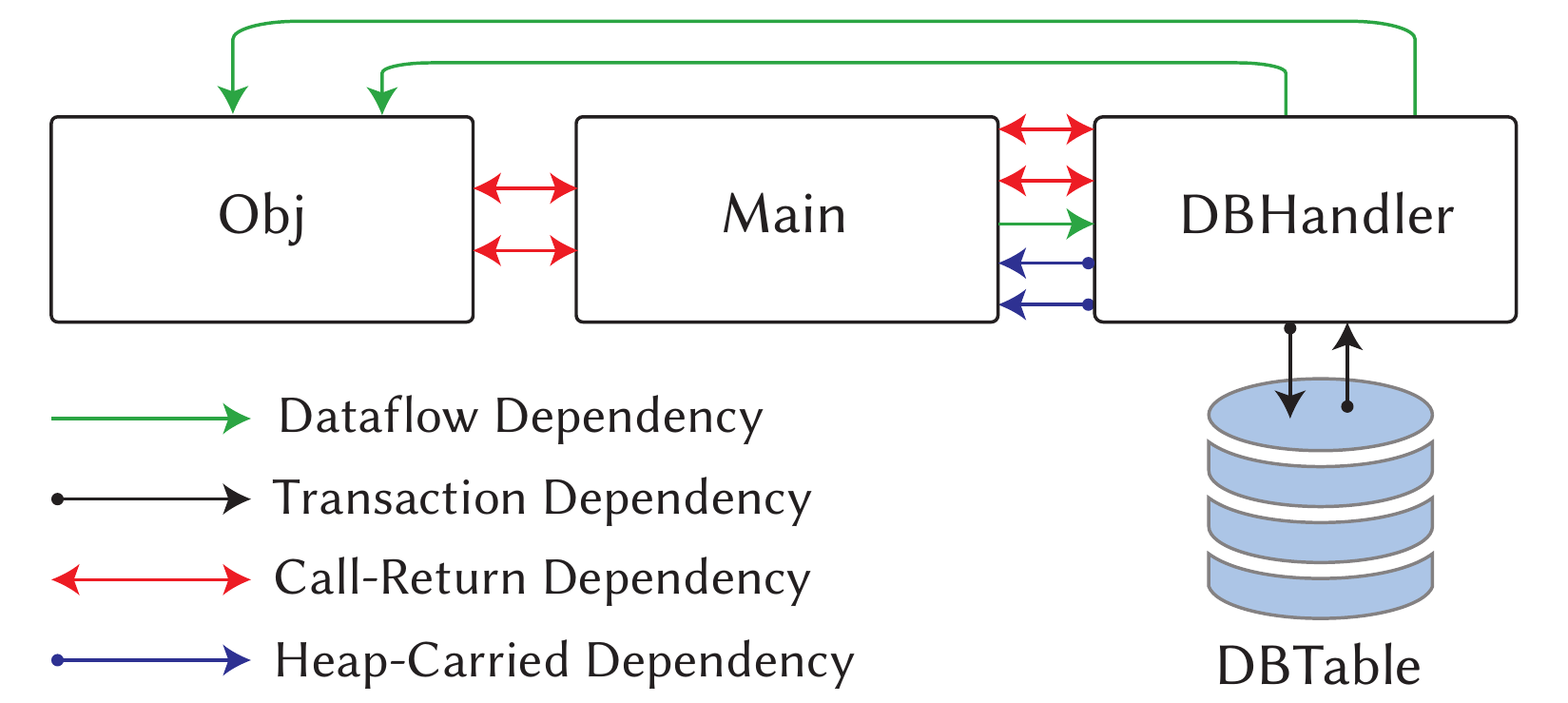}
\caption{A class-level SDG abstraction of the program in \fig{dependencies_a}.}
\label{fig:dependencies_b}
\end{subfigure}
\caption{An example program and it's corresponding SDG. While the program is analyzed at an instruction level granularity, the SDG is constructed at a class level abstraction. Edges in the SDG are implicitly qualified with \texttt{2objH} context-sensitivity. \textit{Note: for the sake of simplicity, this example does not show all the inter-procedural dependencies.}}
\label{fig:dependencies}
\end{figure}

An SDG represents the possible information flows in the form of a graph with nodes (which are classes in our case) and edges representing both the payload of the information as well as the nature of the information flow between methods, \ie, in the form of a call-return relationship, a data-flow relationship, and/or a heap-carried relationship. Further, our formulation of the SDG also highlights the information flow between classes and database tables (which are additional nodes in the SDG) in the form of \textit{transaction reads and writes}. 

\begin{figure*}[h!]
\centering
\includegraphics[width=\linewidth]{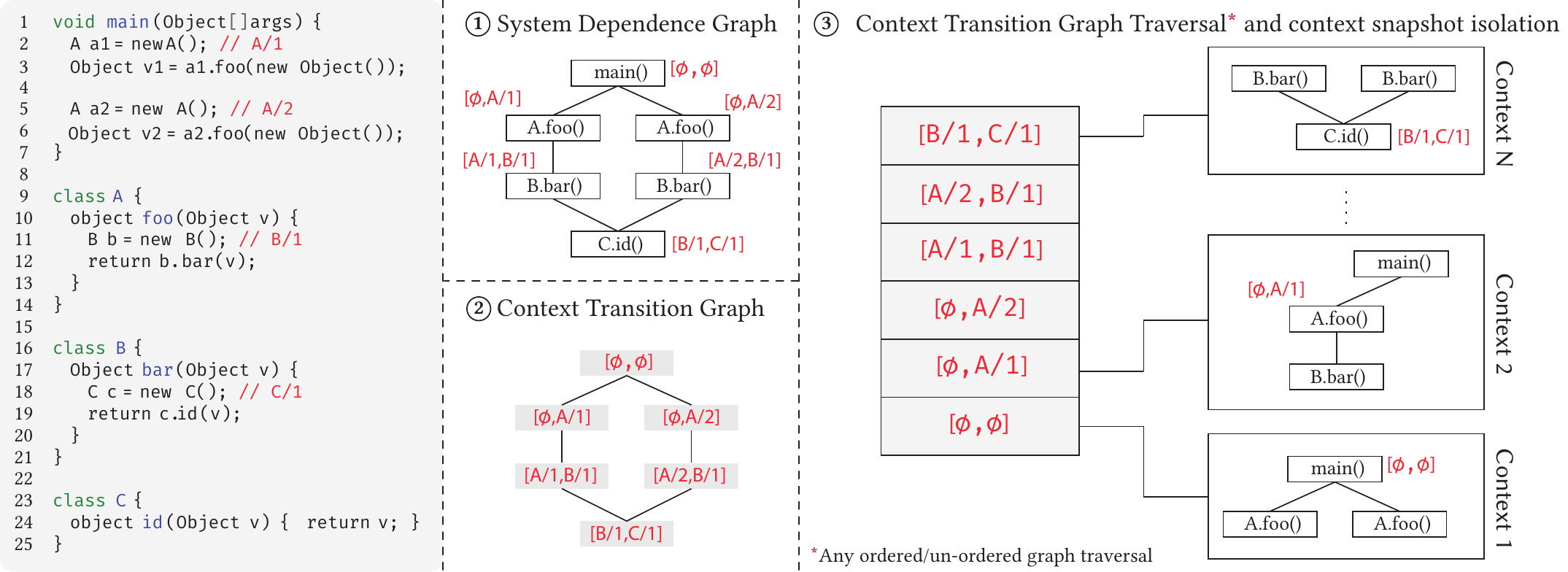}
\caption{A sample workflow depicting the extraction of context-snapshot from the SDG. From left to right, the sample program from \protect\fig{context_sen}; the system dependence graph {\footnotesize\protect\circledw{1}} and context transition graph {\footnotesize\protect\circledw{2}}; SDG sub-graphs under various contexts {\footnotesize\protect\circledw{3}}.}
\label{fig:context_snapshots}
\end{figure*}

\fig{dependencies} illustrates a simple example highlighting the various dependencies that are extracted from the application. Specifically, \fig{dependencies_a} highlights the inter-procedural dependencies between instructions in different classes, where each dependency is annotated with a 2objH context information.  \fig{dependencies_b}, illustrates a class-level abstraction of the SDG in \fig{dependencies_a}. The dependency types are:
\be
\item \textit{Call-Return dependencies:} These are the cornerstone of inter-procedural static analysis. Each edge represents a possible method invocation from an instruction in the source method to a target method. 
Since Java is an object-oriented language, each source and target method has a fully qualified reference object associated with it. This enables us to associate an object-sensitive context with both the caller and the callee.
In \fig{dependencies}, the call-return edges are denoted with  \red{$\longleftrightarrow$}. We note that each call return edge is associated with a \texttt{2ObjH} context, \eg, in ~\fig{dependencies}(a) there are two call-graph edges from class \texttt{Main.main()} to the \texttt{Obj.getVal()} with different contexts \red{\texttt{[$\phi$, \texttt{Obj/1}]}} and \red{\texttt{[$\phi$, \texttt{Obj/2}]}}. 

\item \textit{Data Dependencies:} These edges indicate a static abstraction of the data flows through any two pairs of instructions in the program. Data flow edges (denoted by {\color[RGB]{87,165,81}$\longrightarrow$} in ~\fig{dependencies_a}) can arise in a number of ways: (a) arguments passed through a call-graph edge (\eg, in the call from \texttt{Main.main} to \texttt{DBHandler.write(x,y)}); or (b) variable reads/writes that are propagated across classes (\eg, in the field writes \texttt{o1.val = x)} in \texttt{DBHandler.read()}. It is possible to extract many of these dependencies with taint tracking.

\item \textit{Heap-carried Dependencies:} These edges indicate if two statements access the same object on the heap. In order to detect heap-carried dependencies, we use points-to analysis in order to determine which objects each statement may access. Heap-carried edges (denoted by {\color[RGB]{41,65, 158}$\longrightarrow$} in ~\fig{dependencies_a}) use context-sensitivity to capture the heap-objects with a 1-heap-sensitivity (indicated by the `H' in 2objH). This heap-sensitivity allows us to precisely discover which instances of the heap-objects are being used. A good example of this can be seen in \fig{dependencies_a} where we first write to \texttt{o1.val=x} in \texttt{DBHandler.read()} and then read it back in \texttt{x = o1.getVal()}. This is a typical heap carried dependency that arises due to a shared object {\color[RGB]{41,65, 158} \texttt{[Obj/1]}}. Other heap-carried edges can be differentiated with their corresponding heap-context (\eg, {\color[RGB]{41,65, 158} \texttt{[Obj/2]}} in \fig{dependencies_a}).

\item \textit{Transactions Dependencies:} These edges identify sections of the code that take part in database transactions. Database transactions are a critical and an irreplaceable component of Java EE applications. It is possible to discover transactions using static analyzers like WALA/SOOT/DOOP. To do this, we leverage JEE features such as Java Persistence API (JPA) and JAVA Transaction API (JTA). These APIs allow users to annotate a method with \texttt{@Transactional} to indicate that the method is part of a transaction. We process these methods their instructions to identify the destination database of the transaction.
\ee

We note that, as per the norm~\cite{horwitz1990interprocedural}, we gather the dependencies at an instruction level, but abstract away the intra-partition dependencies to minimize the volume of the graph. Since JEE based monoliths are often partitioned at a class-level, we conjecture that intra-procedural dependencies play a minimal role in enabling that process when compared to inter-procedural dependencies.


\subsection{Partitioning with \CARGO}
\label{sect:stage2}

In this section, we describe \textit{Context sensitive Label Propagation} (or \CARGO)---a novel community detection algorithm that builds upon the principles of Label Propagation Algorithm (LPA)~\cite{xiaojin2002learning} to effectively leverage the rich context-sensitive dependencies in the SDG to identify and demarcate functional boundaries in the code. We first describe off-the-shelf label propagation (\tion{lpa}). Then we discuss motivation for, and construction of, context and transactional snapshots from the SDG (\tion{snapshots}). Finally we describe \CARGO and how label propagation across each snapshot can be used generate highly-cohesive and functionally isolated partitions (\tion{cargo_algo}).

\subsubsection{Label propagation}
\label{sect:lpa}
LPA is a semi-supervised algorithm for identifying communities of densely connected regions in a network that works based just on network structure, without a need for predefined objective functions or a priori knowledge about communities~\cite{xiaojin2002learning}. Our intuition is based on the idea that if sections of the code are cohesive, then they tend to be densely connected and therefore the labels from LPA will propagate quickly through them and a single label will come to dominate that section of the code. Whereas, for sections of the code that are only loosely coupled, labels will fail to propagate to these regions. This essentially isolates loosely coupled regions of the code while coalescing cohesive sections of the code. Consequently, as the Label Propagation converges, each highly cohesive group of nodes (classes) will acquire a consensus on a single label and each loosely coupled regions of the code will have their own unique labels.

\subsubsection{Context and Transactional Snapshots}
\label{sect:snapshots}
Using off-the-shelf label propagation on the entire SDG is inaccurate and may limit us from leveraging the precision and completeness of the SDG. This is because the SDG--in it's entirety--represents a superposition of all the possible contextual states of the program. At any given point, the program may only exist under only one of these contexts (and not all of them simultaneously). Consider the sample program shown in \fig{context_snapshots} and its system dependency graph \circledw{1}. Here, the SDG consists of two versions of the same methods \texttt{A.foo()} and \texttt{B.bar()} under different contexts. At a given instance, these methods can exist under one context, but not both. Therefore, to process the SDG accurately, we need extract sub-graphs of the SDG that represent all the individual program states under various contexts. We denote these sub-graphs with the term ``context snapshots''. 

To extract the context snapshots, we first construct a context transition graph to capture all possible context transitions in the program. This can be derived from the SDG by analyzing all pairs of vertices connected by an edge and recording their contexts. For example, in~\fig{context_snapshots}, we analyze the SDG \circledw{1} to obtain the contexts of each pair of vertices to build the context transition graph seen in \circledw{2}. Next, for each context (say $ctx_i$) in the context transition graph, we extract all the nodes from the SDG belonging to the current context $ctx_i$ as well as the nodes belonging to the preceding context ($ctx_{i-1}$) and the succeeding context ($ctx_{i+1}$). The key intuition here is that, for any given context, we must be able to know three things: (a) what is the current state (\ie, $ctx_i$)?, (b) how did we get to the current state (\ie, the previous contexts $ctx_{i-1}$)?, and (c)~which program state can we transition to (\ie, the next contexts $ctx_{i+1}$)? Together, these nodes represent a ``snapshot'' of the program's state. This is repeated for all contexts to get a list of ``context-snapshots''.

In a spirit similar to that of context-snapshots, we also generate a \textit{transactional snapshot} which is a sub-graph of the SDG that includes vertices that are databases and those classes which engage in transactions with this database.
It produces a perspective of the application that is centered explicitly around databases. Our key intuition for including this view is to encourage our label propagation to explicitly collect classes that share transactional dependencies with one another. This should, in theory, reduce incidences of distributed transactions.

\subsubsection{Context Sensitive Label Propagation (\CARGO)}
\label{sect:cargo_algo}
Context Sensitivity Label Propagation (\CARGO) is a novel variant of LPA that attempts to leverage the SDG (particularly, the context- and transactional-snapshots) to identify functionally similar and loosely coupled components of the application. It is comprised of 3 steps, namely (1) Initialization with seed labels; (2) Initial LPA with the transactional snapshot; and (3) Iterative LPA over context snapshots. Each of these are described in detail below:
\bi
\item[$\mathbf{\circ}$]\textbf{Step 1:~}\textit{Initialization with seed labels.}~Label propagation, and by extension \CARGO, are required to be seeded with initial partitioning labels. In the presence of an existing partitioning assignment via the use of an existing algorithm (or a user-preferred partitioning assignment), we may use that to seed \CARGO. In the absence of these seeds, we assign each class its own partition ID. In this sense, \CARGO can be fully unsupervised, or it may be semi-supervised.
\item[$\mathbf{\circ}$]\textbf{Step 2:~}\textit{Initial LPA with the transactional snapshot.} In the first stage, we apply LPA over the transactional snapshot of the SDG. This identifies classes that engage in distributed transactions and propagates seed labels between them. If more than one class shares transaction edges with the same DB table, then LPA attempts to assign the same label to them. For two or more classes that do not write to the same table, their labels will remain unchanged. While LPA doesn't guarantee that distributed transactions will be eliminated, this initial seeding reduces incidents of distributed transactions in the subsequent stages as other class nodes adapt to the seeds from this stage.
\item[$\mathbf{\circ}$]\textbf{Step 3:~}\textit{Iterative LPA over context-snapshots.} In this stage, we iteratively apply label propagation on each context snapshot of the SDG. The underlying intuition is as follows---in a significant number of snapshots, if a group of classes seem to be connected to one another (through call, data, or heap-dependency), they are likely cohesive and must thus belong together. In contrast, if classes are seldom connected across all context snapshots, they are weakly tied and must exist in different partitions. As we apply LPA to each context snapshot, classes that are cohesive, frequently co-occurring, and related will quickly share the same label. For classes that do not frequently co-occur, a separate label will be used. This means that \CARGO can isolate weakly connected portions of the code while combining cohesive portions of the code in a single partition. 
\ei

\section{Evaluation}
\label{sec:evaluation}

\subsection{Benchmark Applications}

We use 5 commonly used Java applications as our benchmarks to perform evaluation. These include: (1) Daytrader~\cite{daytrader-app}: A Java EE7 application built around the paradigm of an online stock trading system. The application allows users to login, view their portfolio, and buy or sell stock shares; (2) Plants~\cite{plants-app}: a simple Java EE 6 application which uses CDI managed beans, Java Server Faces (JSF), and Java Server Pages (JSP); (3) AcmeAir \cite{acmeair-app}: a Java web application for a fictious airline company which allows users to book and manage flights; JPetStore \cite{jpetstore-app}: a Java web application for a pet store where users shop for pets online; (5) one proprietary app which we call Proprietary1. They contain 109, 33, 66, 37 and 82 classes respectively. We note that of the above benchmark applications, Daytrader uses 6 SQL database tables for its business logic.


\subsection{Implementation}
We use Doop \cite{bravenboer2009strictly} and datalog to implement all our static analysis as described in \tion{methods}. We also leverage JackEE \cite{antoniadis2020static}, a framework that performs scalable and precise analysis of Java Enterprise applications. Our datalog rules will be made available as part of our published code. We then use \texttt{networkx} \cite{SciPyProceedings_11} to parse the generated edges and build an SDG that represents the program. We run \CARGO in two modes:
\be
\item \textbf{Native mode:} This is a purely unsupervised version of the algorithm, where we are provided only with a maximum number of partitions $K$. We take a random ordering of the  nodes of the SDG, and initialize the $i^{\text{th}}$ node with the label $(i\;\%\;K)$. We then run \CARGO as described in \tion{methods} using these initial labels to arrive at partition assignments for each node.

\item \textbf{Refinement mode:} This is a semi-supervised version of the algorithm, where we are provided with an initial set of seed labels obtained from another partitioning approach. These labels are assigned to the corresponding nodes of the SDG, and the rest of the SDG nodes are initialized with "-1", denoting no label. We then run \CARGO to propagate these initial labels. When the algorithm finishes, if nodes are still labeled "-1", these are unreachable classes.
\ee

\noindent\textbf{Note:} In the rest of the paper, each algorithm \textsc{Algorithm} refined with \CARGO is denoted by  \textsc{Algorithm++}. The name \CARGO refers to the algorithm run in native mode, unless specified otherwise.

\subsection{Metrics}
\label{sect:metrics}
\subsubsection{Database Transactional Purity}

To minimize distributed transactions, we would like each database object to be accessed by classes from one partition only. To design a metric that quantifies this, we use entropy~\cite{shannon2001mathematical} which measures the homogeneity of a set of samples. A low value of entropy indicates a high purity of the samples and vice-versa.

For each database table, find the list of partitions from which it is accessed. Let the entropy of this list be $h$; then transactional purity for this database table is defined as $1-h$. If a database table is accessed only by classes from a single partition, the transactional purity for that table will be $1.0$. The transactional purity of the whole program is the average of the transactional purity across all database tables. Higher transactional purity is indicative of fewer distributed transactions, which is a desirable property for a partitioning scheme.

\subsubsection{Run-time Performance Metrics}
These measures evaluate the run-time behavior of the program.
\bi
    \item \textbf{Latency} measures how long, in milliseconds, it takes to perform a user action. We calculate average latency of all the requests made by all the users.
    \item \textbf{Throughput} measures how many requests are completed per second, averaged over the entire run time of the program.
\ei

\subsubsection{Architectural Metrics}
Using our PDG and dynamic call graphs, we define four metrics to measure the effectiveness of partitions:

\begin{itemize}[wide=0pt]
\item \textbf{Coupling} \cite{martin2003agile} measures the degree of interaction between different partitions of the microservice. Coupling for each partition is the number of edges in the SDG going from or to the partition. We calculate the sum of coupling across all partitions, and divide this by the total number of edges in the SDG.

\item \textbf{Cohesion} \cite{martin2003agile} measures the degree of interconnectedness between classes in a partition. For each partition, it is computed as $\texttt{(internal edges)}/ (\texttt{internal edges} + \texttt{external edges})$. We calculate cohesion for each partition separately and compute the mean across all partitions.

\item \textbf{Inter-call Percentage (ICP)} \cite{kalia2021mono2micro} measures the percentage of the overall call volume occurring between two partitions. $\texttt{ICP}_{i, j} = c_{i,j} / \sum_{i \neq j}^{M} c_{i,j}$, where $c_{i,j}$ represents the number of calls between partition $i$ and partition $j$. Lower the ICP, better the partitioning.

\item \textbf{Business Context Purity (BCP)} \cite{Kalia+2020+FSE} measures the average entropy of business use cases per partition. Each class is associated with one use case, so each partition has a list of use cases. To maintain semantic and functional coherence of the partitions, we would like each partition to not handle too many different use cases. A list consisting of many different kinds of use cases would have high entropy and high BCP, so lower BCP means better partitions.

\end{itemize}
\section{Experimental Results}

We evaluate our approach through the following questions :
\bi
    \item \textbf{RQ1: Mitigating distributed Transactions.} How effective is our approach in  minimizing distributed transactions?
    
    \item \textbf{RQ2: Run-time performance.} How do the latency and throughput of a refined microservice application compare to those of the original application?
    
	\item \textbf{RQ3: Performance on architectural metrics.} How does our approach perform on architectural metrics?
\ei

\subsection{RQ1: Distributed Database Transactions}
Mitigating distributed transactions is crucial for a microservice application. In this section we evaluate how effective our approach is in minimizing distributed transactions.

\noindent\textbf{\textit{Evaluation:}} We run \CARGO to refine the output of other partitioning algorithms. We also run \CARGO stand-alone. Since Daytrader is the only dataset that uses external databases, we run all partitioning algorithms only on Daytrader. We run each algorithm with $K = [3, 5, 7, 9, 11, 13]$ and 5 random seeds, and measure the average transactional purity across all seeds and values of $K$.

\noindent\textbf{\textit{Discussion:}} The transactional purity is shown in~\fig{rq1_db}. \CARGO increases the transactional purity of all baseline approaches (\ie, approaches suffixed with $++$ perform better than the ones without the suffix). When used in native mode (\CARGO), and when used to refine the outputs of FoSCI (FoSCI++) and MEM (MEM++), \CARGO achieves transactional purity of $1.0$. This indicates that there is no database table that is accessed by classes from multiple partitions.
\\\\
\begin{result}
    For each of the 4 baselines, the refined partitions have higher transactional purity. Further, FoSCI++, MEM++ and \CARGO have transactional purity of 1.0, implying that there can be no distributed transactions with these partitioning schemes.
\end{result}

\subsection{RQ2 : Run-time Performance}
\begin{figure}[t!]
    \centering
    \includegraphics[width=0.85\linewidth]{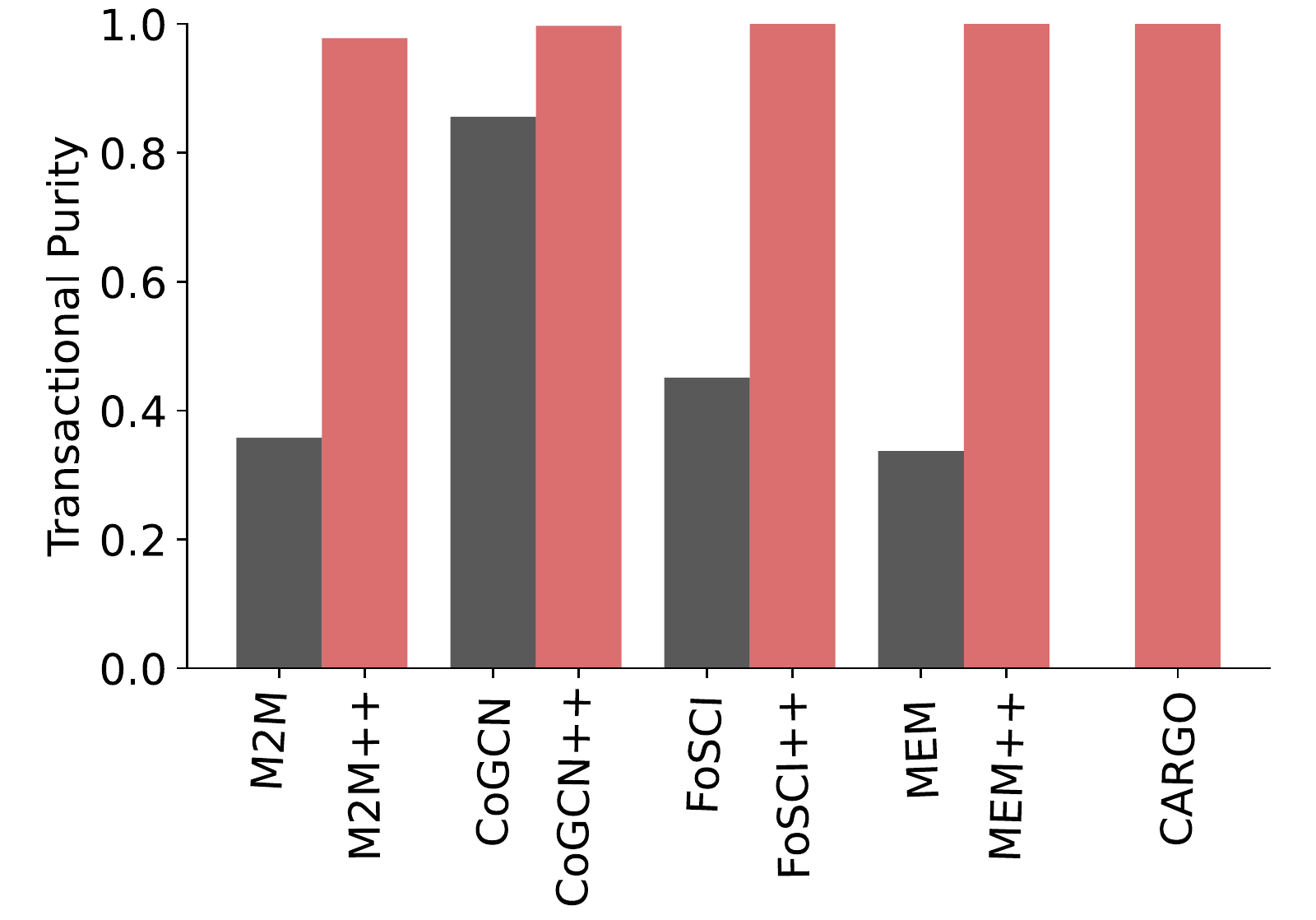}
    \caption{Transaction purity for \CARGO and other approaches. The transactional purity of the refined partitions is higher than that of the original partitions for all approaches. FoSCI++, MEM++ and \CARGO achieve transactional purity of 1.0, indicating that no database objects is accessed by classes belonging to different partitions.}
    \label{fig:rq1_db}
\end{figure}
\begin{figure*}[t!]
    \centering
    \includegraphics[width=\linewidth]{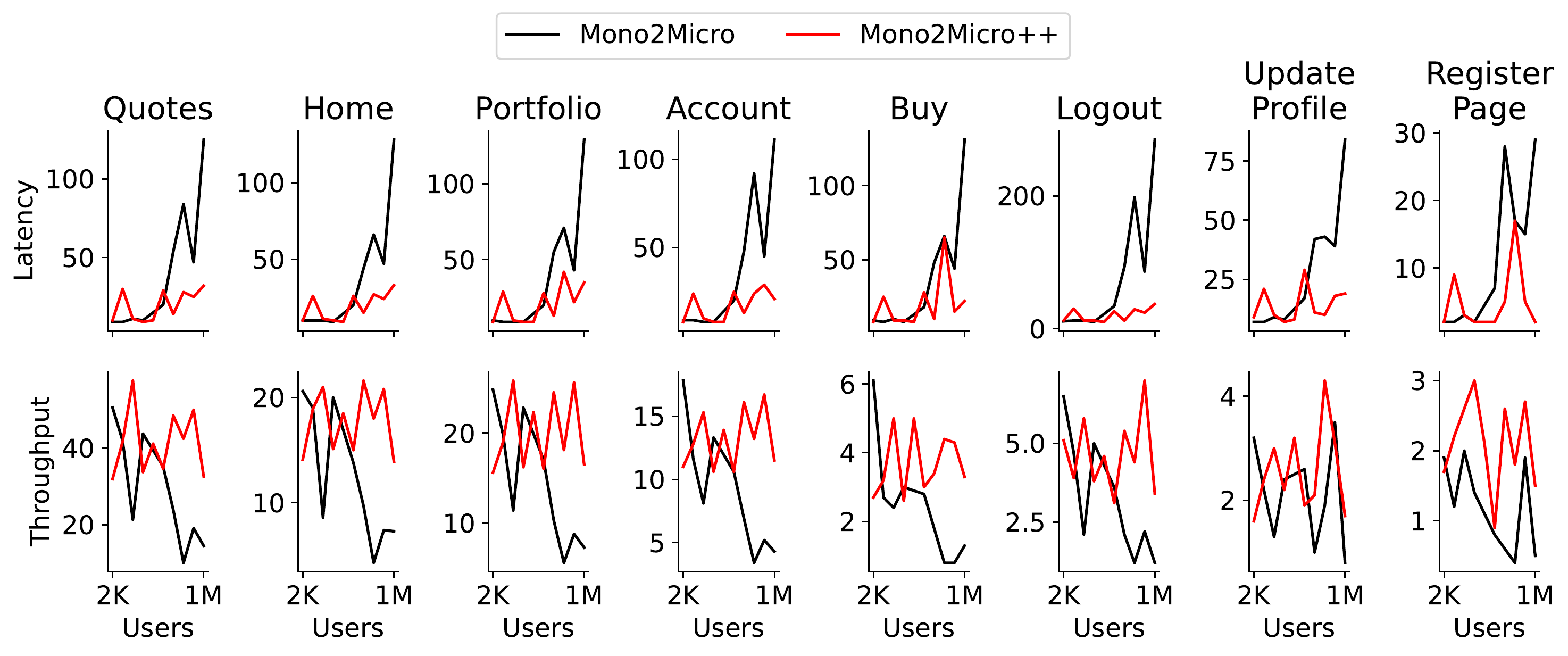}
        \caption{Latency and Throughput for Mono2Micro versus Mono2Micro++, for different use cases in Daytrader. We simulate each application with the number of users varying from $2^{11} \approx 2K$ to $2^{20} \approx 1M$ in exponents of $2$. Mono2Micro++ achieves lower latency and higher throughput than Mono2Micro, indicating better performance.}
        \label{fig:latency}
\end{figure*}
\setlength{\tabcolsep}{2.5pt}

\begin{table*}[t!]

    \caption{Performance of different approaches on four architectural metrics computed on 5 applications. We evaluate Mono2Micro (M2M), CoGCN, FoSCI and MEM, along with their refined partitions denoted with a "++". We also evaluate \CARGO stand-alone. The symbol $\triangle$ next to a metric indicates that higher is better, while $\triangledown$ indicates that lower is better. For each approach and refined approach, we shade the better of the two \colorbox{blue!10}{blue}. We count the number of wins, ties and losses for the refined approaches versus the original approaches across all the datasets, and shade the corresponding cell \colorbox{green!10}{green} if wins outnumber the losses. The best performing approach in each row is \textbf{bold}. If \CARGO is the best performing approach, then we shade its cell \colorbox{orange!10}{orange}}

\resizebox{0.9\linewidth}{!}{
    \begin{tabular}{|r|rr|rr|rr|rr|c||rr|rr|rr|rr|c|}
        \cline{2-19}
         \multicolumn{1}{c|}{} & \multicolumn{9}{c||}{\textbf{Coupling $\triangledown$}} & \multicolumn{9}{c|}{\textbf{Cohesion $\triangle$}} \bigstrut\\
        \cline{2-19}
        \multicolumn{1}{c|}{} &
        \rot{90}{\textbf{M2M}} &
        \rot{90}{\textbf{M2M++}} &
        \rot{90}{\textbf{CoGCN}} &
        \rot{90}{\textbf{CoGCN++\hspace{4pt}}} &
        \rot{90}{\textbf{FoSCI}} &
        \rot{90}{\textbf{FoSCI++}} &
        \rot{90}{\textbf{MEM}} &
        \rot{90}{\textbf{MEM++}} &
        \rot{90}{\textbf{\CARGO}} &
        \rot{90}{\textbf{M2M}} &
        \rot{90}{\textbf{M2M++}} &
        \rot{90}{\textbf{CoGCN}} &
        \rot{90}{\textbf{CoGCN++\hspace{4pt}}} &
        \rot{90}{\textbf{FoSCI}} &
        \rot{90}{\textbf{FoSCI++}} &
        \rot{90}{\textbf{MEM}} &
        \rot{90}{\textbf{MEM++}} &
        \rot{90}{\textbf{\CARGO}}\bigstrut\\
         \cline{2-19} \noalign{\vskip\doublerulesep
         \vskip-\arrayrulewidth}
         \hline
		 \textsc{Daytrader} & 0.78 & \cb{0.02} & 0.25 & \cb{0.02} & 0.68 & \cb{0.02} & 0.12 & \cb{\bfseries 0.01} & \bfseries \cp{0.01} & 0.30 & \cb{0.62} & 0.37 & \cb{\bfseries 0.61} & 0.23 & \cb{0.54} & 0.26 & \cb{0.46} & \cp{\bfseries 0.71}\bigstrut\\

		 \textsc{Plants} & 0.31 & \cb{\bfseries 0.04} & 0.40 & \cb{0.29} & 0.75 & \cb{0.31} & 0.29 & \cb{0.11} & 0.05 & 0.32 & \cb{\bfseries 0.61} & 0.39 & \cb{0.46} & 0.24 & \cb{0.40} & 0.39 & \cb{0.44} & 0.6\bigstrut\\

		 \textsc{Acmeair} & 0.58 & \cb{0.04} & 0.52 & \cb{0.13} & 0.71 & \cb{0.22} & 0.51 & \cb{0.15} & \cp{\bfseries 0.03} & 0.27 & \cb{0.48} & 0.21 & \cb{0.32} & 0.16 & \cb{0.51} & 0.37 & \cb{0.55} & 
		 \cp{\bfseries 0.96}\bigstrut\\

		 \textsc{Jpetstore} & 0.77 & \cb{\bfseries 0.03} & 0.78 & \cb{0.06} & 0.79 & \cb{\bfseries 0.03} & 0.77 & \cb{0.05} & \bfseries 0.03 & 0.27 & \cb{0.29} & 0.20 & \cb{0.24} & 0.18 & \cb{0.36} & 0.30 & \cb{0.44} & \cp{\bfseries 0.94}\bigstrut\\

		 \textsc{Proprietary1} & 0.42 & \cb{\bfseries 0.03} & 0.22 & \cb{0.04} & 0.48 & \cb{0.02} & 0.27 & \cb{\bfseries 0.03} & 0.04 & 0.57 & \cb{\bfseries 0.78} & 0.69 & \cb{0.73} & 0.37 & \cb{0.57} & 0.48 & \cb{0.67} & 0.75\bigstrut\\
         \hline
         \hline
         \textsc{Win/Tie/Loss} & \multicolumn{2}{c|}{\cg{5 / 0 / 0}} & \multicolumn{2}{c|}{\cg{5 / 0 / 0}} & \multicolumn{2}{c|}{\cg{5 / 0 / 0}} & \multicolumn{2}{c|}{\cg{5 / 0 / 0}} &  & \multicolumn{2}{c|}{\cg{5 / 0 / 0}} & \multicolumn{2}{c|}{\cg{5 / 0 / 0}} & \multicolumn{2}{c|}{\cg{5 / 0 / 0}} & \multicolumn{2}{c|}{\cg{5 / 0 / 0}} & \\
         \hline
    \end{tabular}}
    \\[1em]
    \setlength{\tabcolsep}{2pt}
    \resizebox{0.9\linewidth}{!}{\begin{tabular}{|r|rr|rr|rr|rr|c||rr|rr|rr|rr|c|}
        \cline{2-19}
         \multicolumn{1}{c|}{} & \multicolumn{9}{c||}{\textbf{BCP $\triangledown$}} & \multicolumn{9}{c|}{\textbf{ICP $\triangledown$}} \bigstrut\\
        \cline{2-19}
        \multicolumn{1}{c|}{} &
        \rot{90}{\textbf{M2M}} &
        \rot{90}{\textbf{M2M++}} &
        \rot{90}{\textbf{CoGCN}} &
        \rot{90}{\textbf{CoGCN++\hspace{4pt}}} &
        \rot{90}{\textbf{FoSCI}} &
        \rot{90}{\textbf{FoSCI++}} &
        \rot{90}{\textbf{MEM}} &
        \rot{90}{\textbf{MEM++}} &
        \rot{90}{\textbf{\CARGO}} &
        \rot{90}{\textbf{M2M}} &
        \rot{90}{\textbf{M2M++}} &
        \rot{90}{\textbf{CoGCN}} &
        \rot{90}{\textbf{CoGCN++\hspace{4pt}}} &
        \rot{90}{\textbf{FoSCI}} &
        \rot{90}{\textbf{FoSCI++}} &
        \rot{90}{\textbf{MEM}} &
        \rot{90}{\textbf{MEM++}} &
        \rot{90}{\textbf{\CARGO}}
        \bigstrut\\
         \cline{2-19} \noalign{\vskip\doublerulesep
         \vskip-\arrayrulewidth}
         \hline
		\textsc{Daytrader} & \cb{2.31} & 2.57 & \cb{1.40} & 1.90 & 2.55 & \cb{2.26} & \cb{2.23} & 2.32 & \cp{\bfseries 1.31} & 0.60 & \cb{0.05} & 0.16 & \cb{0.05} & 0.48 & \cb{0.07} & \cb{0.04} & 0.07 & \cp{\bfseries 0.00}\bigstrut\\

		\textsc{Plants} & \cb{1.68} & 2.20 & 1.96 & \cb{1.54} & 2.37 & \cb{1.71} & 1.93 & \cb{\bfseries 1.53} & 1.79 & 0.54 & \cb{0.11} & 0.65 & \cb{0.11} & 0.70 & \cb{0.12} & 0.28 & \cb{0.11} & \cp{\bfseries 0.06}\bigstrut\\

		\textsc{Acmeair} & \cb{1.29} & 1.48 & 1.18 & \cb{\bfseries 1.12} & 1.75 & \cb{1.61} & \cb{1.94} & 1.95 & 1.75 & 0.91 & \cb{0.35} & 0.67 & \cb{0.63} & 0.85 & \cb{0.63} & \cb{0.41} & 0.72 & \cp{\bfseries 0.32}\bigstrut\\

		\textsc{Jpetstore} & \cb{2.25} & 2.35 & \cb{\bfseries 1.71} & 1.77 & \cb{2.26} & 2.45 & \cb{2.54} & 2.74 & 2.87 & \cb{\bfseries 0.13} & 0.41 & \cb{0.33} & 0.35 & \cb{0.24} & 0.40 & \cb{0.26} & 0.42 & 0.30\bigstrut\\

		\textsc{Proprietary1} & \cb{\bfseries 1.23} & 1.53 & \cb{1.27} & 1.42 & 1.49 & \cb{1.48} & \cb{1.48} & 1.95 & 1.55 & \cb{\bfseries 0.02} & 0.35 & \cb{0.17} & 0.25 & 0.37 & \cb{0.32} & 0.32 & \cb{0.32} & 0.37\bigstrut\\

		\hline

		\hline

		\textsc{{Win/Tie/Loss}} & \multicolumn{2}{c|}{0 / 0 / 5} & \multicolumn{2}{c|}{2 / 0 / 3} & \multicolumn{2}{c|}{\cg{4 / 0 / 1}} & \multicolumn{2}{c|}{1 / 0 / 4} & & \multicolumn{2}{c|}{\cg{3 / 0 / 2}} & \multicolumn{2}{c|}{\cg{3 / 0 / 2}} & \multicolumn{2}{c|}{\cg{4 / 0 / 1}} & \multicolumn{2}{c|}{1 / 1 / 3} & \\
         \hline
    \end{tabular}}
    \label{table:metrics}
\end{table*}
In this section, we evaluate the performance of our partitioning algorithm in a real world setting. When we implement the recommended partitioning scheme as a microservice application, we want the application to have \textbf{low latency} and \textbf{high throughput}.

\noindent\textbf{\textit{Evaluation:}} The Daytrader application has the largest number of classes among our benchmark applications, so we choose it for these performance evaluations. Daytrader is also the only benchmark to use external databases.

We use Mono2Micro \cite{m2m_blog} to construct a microservice version of the monolithic Daytrader application, and repeat this process for Mono2Micro++. Both these apps are deployed by using individual Docker containers for each partition, and orchestrated using \texttt{docker-compose}.

We use Apache JMeter to measure latency and throughput under varying loads. We simulate various use cases consisting of sequences of user actions. We test the application with multiple users executing concurrent requests every second. The number of users is varied from $2^{11} \approx 2K$ to $2^{20} \approx 1M$ in exponents of $2$.


\textbf{\textit{Discussion:}} The latency and throughput are plotted as a function of the number of users, for different use cases. The results are shown in Figure \ref{fig:latency}. We observe that:
\bi
\item When the number of users is sufficiently high, the Mono2Micro++ partitioning scheme achieves lower latency and higher throughput than the original Mono2Micro partitioning scheme. As the load on the application increases, the coupling between microservices becomes more of an impediment to performance. We hypothesize that Mono2Micro++ performs better than Mono2Micro under high loads because it has less inter-partition coupling.
\item This holds true across all the use cases that we have considered. 
\item We compute the average percentage improvement in latency and throughput across all use cases, and find that Mono2Micro++ represents a 11\% improvement in latency and a 120\% improvement throughput, over Mono2Micro.
\ei
~\\
\begin{result}
When we use our approach to re-partition a microservice application, the resulting application has \textbf{11\%} lower latency and \textbf{120\%} higher throughput on average across use cases than the original application.
\end{result}

\subsection{RQ3: Performance on Architectural Metrics}

RQ1 measured how effectively \CARGO minimizes distributed transactions, and RQ2 evaluated the real-world performance of \CARGO. In this section, we evaluate \CARGO on architectural metrics that use PDG and call-graph information to measure the quality of partitions. We would like to see loose coupling between partitions, and high cohesion within classes of the same partition.

\noindent\textbf{\textit{Evaluation:}} We use four baseline partitioning approaches - \\Mono2Micro (M2M) \cite{kalia2021mono2micro}, CoGCN \cite{Desai+AAAI+2021}, FoSCI 
\cite{Jin+TSE+2019} and MEM \cite{Mazlami+IEEE+2017}. We run these approaches on our benchmark set of five monolithic applications. Each of these approaches allows one to specify a maximum number of partitions, $K$. We run each algorithm with $K = [3, 5, 7, 9, 11, 13]$, and use \CARGO to refine the partitions produced for each value of $K$. Additionally, we run \CARGO in native mode with the same values of $K$. Both when refining existing partitions and while running \CARGO stand-alone, we run the algorithm five times with different random seeds. For each metric, we report the average across all seeds and values of $K$.

\noindent\textbf{\textit{Discussion:}} The results are shown in Table \ref{table:metrics}. The best results are highlighted in \colorbox{blue!10}{blue} We make the following observations:

\bi

\item[$\bullet$] \textit{Cohesion and Coupling:} When used to refine the partitions produced by another algorithm, \CARGO always improves the Cohesion and Coupling. Further, when used stand-alone, \CARGO achieves the best Cohesion and Coupling across all algorithms.

\item[$\bullet$] \textit{ICP:} When used to refine the partitions produced by another algorithm, the refined partitions are noticeably better than the original partitions in terms of ICP. When used in native mode, \CARGO performs significantly better than all the other approaches in 3 out of 5 datasets.

\item[$\bullet$] \textit{BCP:} BCP measures the semantic coherence of each partition in terms of business use cases that it handles. When \CARGO is used stand-alone on Daytrader, it outperforms all other approaches on this metric. However when used to refine existing partitions, it performs poorly. The definition of BCP depends heavily on the quality of the generated business use cases for each class, which is information that \CARGO does not have access to. This is one possible explanation for its poor performance on this metric.
\ei

Finally, we note that \CARGO when used stand-alone with randomly initialized labels achieves state-of-the-art performance on all four metrics on Daytrader. This shows that even in the absence of domain knowledge, \CARGO can effectively learn the structure of the program and form partitions.
\\\\
\begin{result}
In 3 out of the 4 metrics, the refined partitions are better than the original partitions for a majority of the approaches. Further, \CARGO used in native mode achieves state-of-the-art performance on all 4 metrics on the Daytrader benchmark.
\end{result}

\section{Related Work}

\textbf{Software Modularization:} Microservice decomposition is a specific instance of the Software Modularization problem, which has seen a long line of work. In general, software modularization approaches construct a Module Dependence Graph (MDG), and perform various analyses on this graph. \citet{doval1999automatic}, \citet{mitchell2006automatic} and \citet{harman2002new} define a modularity metric based on intra-cluster and inter-cluster connections and optimize this metric using genetic algorithms or hill climbing. \citet{sarkar2006api} define information-theory based metrics to evaluate the quality of software modularization. \citet{bavota2013using} quantify the cohesion between classes using the volume of information flowing between functions, and the semantic similarity of the identifiers and comments used.

\noindent\textbf{Decomposition into Microservices:} There have been several recent papers that attempt to decompose monolithic applications into microservices \cite{ahmadvand2016requirements, baresi2017microservices, chen2017monolith, escobar2016towards, gysel2016service, hassan2017microservice, klock2017workload, levcovitz2016towards, mazlami2017extraction}. \citet{escobar2016towards} present a rule-based algorithm to partition JEE applications that separates classes according to the data that they manipulate. \cite{mazlami2017extraction} define 3 coupling criteria between classes based on an analysis of commit patterns and semantic similarity. They then construct a graph where the weight of an edge connecting two classes is proportional to the strength of the coupling between them, and use clustering algorithms on this graph. \citet{gysel2016service} propose a similar algorithm, with 16 coupling criteria instead of 3.
There are also some approaches that use dynamic traces to cluster classes into microservices. \citet{jin2018functionality} identify functional atoms, or groups of classes that frequently occur together in a trace, and define four optimization objectives to group these functional atoms. \citet{desai2021graph} construct a co-occurrence graph between classes that occur in the same execution trace, and use a Graph Neural Network (GNN) to obtain node embeddings and cluster nodes on this graph. \cite{kalia2021mono2micro} measure the number of direct and indirect calls between classes in a dynamic trace, and apply hierarchical clustering on the similarity graph thus obtained.
\section{Threats to Validity}
As with any evaluation study, biases can affect the final results. Therefore, any conclusions of this work must be considered with the following issues in mind:

$\bullet$ \textit{Evaluation Bias}: 
In  {\bf RQ1} and {\bf RQ3}, we compare the performance of \CARGO using a set of architectural metrics (defined in \tion{metrics}). The results and the conclusions of these RQs are scoped by the efficacy of the metrics to evaluate the architecture. Since these metrics are statically computed, there may be slightly different conclusions with other measurements. This is to be explored in future research.  

$\bullet$ \textit{Construct Validity}: At various places in this paper, we made engineering decisions about program graph granularity, traversal orders, and so on. While these decisions were made using advice from the literature, we acknowledge that other constructs might lead to other conclusions. To enable further exploration of our design decisions and in the interest of open science, we make available our replication package.

$\bullet$ \textit{External Validity}: For this study, we have selected a diverse set of subject systems, including a proprietary application. We also use a number of microservice partitioning algorithms for comparison. There is a possibility that our outcomes may vary for a different set of applications. Therefore, one has to be careful when generalizing our findings to other applications or algorithms. Even though we tried to run our experiment on a variety of applications, we do not claim that our results generalize beyond the specific case studies explored here. That said, to enable reproducibility, we have shared our scripts and the gathered performance data. 

\section{Conclusion}

Microservices Architecture (MSA) has evolved into a de-facto standard for developing cloud-native enterprise applications during the last decade. A number of automated approaches are being developed to recommend functional partitions in monolithic code that can be extracted away and modernized as microservices. However, many of these approaches produce recommendations that are not actionable as they result in (a)~distributed monoliths, and/or (b)~force the use of distributed transactions. This work attempts to overcome these challenges by introducing CARGO-—a novel un-/semi-supervised partition refinement technique that was used to refine and thereby enrich the partitioning quality of the current state-of-the-art algorithms. Our findings over several Java EE projects suggest that \CARGO can improve the partition quality of the existing approach and steer the recommendations towards generating better microservice recommendations.

\bibliographystyle{ACM-Reference-Format}
\bibliography{references}

\end{document}